\documentclass[prl,twocolumn,superscriptaddress, floatfix]{revtex4-2}

\usepackage{graphicx,color}
\usepackage{amssymb}
\usepackage{amsmath}
\usepackage{amsfonts,amsthm} 
\usepackage{natbib}
\usepackage[colorlinks=true,linkcolor=blue,citecolor=blue,urlcolor=blue]{hyperref}
\usepackage[normalem]{ulem}
\usepackage{physics}
\usepackage{placeins}

\newcommand{\HTO}{Ho$_2$Ti$_2$O$_7$}

\newcommand{\HIO}{Ho$_2$Ir$_2$O$_7$}

\newcommand{\HRO}{Ho$_2$Ru$_2$O$_7$}
\newcommand{\LRO}{Lu$_2$Ru$_2$O$_7$}

\begin{document}

\title{Ferromagnetic fragmented state in the pyrochlore \HRO}

\author{F. Museur}
\email[Contact: ]{flavien.museur@psi.ch}
\altaffiliation[Current address: ]{Laboratory for Mesoscopic Systems, Paul Scherrer Institut and ETH Zurich, 8093 Zurich, Switzerland.}
\affiliation{Institut N\'eel, CNRS \& Univ. Grenoble Alpes, 38042 Grenoble, France}
\affiliation{ENS de Lyon, CNRS, Laboratoire de Physique, F-69342 Lyon, France}
\author{J. Robert}
\affiliation{Institut N\'eel, CNRS \& Univ. Grenoble Alpes, 38042 Grenoble, France}
\author{F. Morineau}
\affiliation{Institut N\'eel, CNRS \& Univ. Grenoble Alpes, 38042 Grenoble, France}
\author{N. Bujault}
\affiliation{Institut N\'eel, CNRS \& Univ. Grenoble Alpes, 38042 Grenoble, France}
\author{V. Simonet}
\affiliation{Institut N\'eel, CNRS \& Univ. Grenoble Alpes, 38042 Grenoble, France}
\author{E. Pachoud}
\affiliation{Institut N\'eel, CNRS \& Univ. Grenoble Alpes, 38042 Grenoble, France}
\author{A. Hadj-Azzem}
\affiliation{Institut N\'eel, CNRS \& Univ. Grenoble Alpes, 38042 Grenoble, France}
\author{C. Colin}
\affiliation{Institut N\'eel, CNRS \& Univ. Grenoble Alpes, 38042 Grenoble, France}
\author{L. Mangin-Thro}
\affiliation{Institut Laue Langevin, 71 avenue des Martyrs, CS 20156, 38042 Grenoble Cedex 9, France}
\author{P. Manuel}
\affiliation{ISIS Facility, Rutherford Appleton Laboratory-STFC, Chilton, Didcot, OX11 0QX, United Kingdom}
\author{J. R. Stewart}
\affiliation{ISIS Facility, Rutherford Appleton Laboratory-STFC, Chilton, Didcot, OX11 0QX, United Kingdom}
\author{P. C. W. Holdsworth}
\affiliation{ENS de Lyon, CNRS, Laboratoire de Physique, F-69342 Lyon, France}
\affiliation{French American Center for Theoretical Science, CNRS, KITP, Santa Barbara, CA 93106-4030, USA}
\author{E. Lhotel}
\affiliation{Institut N\'eel, CNRS \& Univ. Grenoble Alpes, 38042 Grenoble, France}

\begin{abstract}
The consecutive magnetic ordering of the Ho and Ru ions in the pyrochlore \HRO\ and their interplay are investigated by neutron scattering, magnetic and specific heat measurements. The Ru moments order at 95 K into a $\Gamma_5$ easy-plane antiferromagnetic state. At 1.55 K the Ho moments order into an unusual $\Gamma_9$ ferromagnetic state with extensive ground state entropy and structured spin dynamics. It is shown how  the internal fields with  $\Gamma_5$ and $\Gamma_9$ geometry allow for two symmetry breaking transitions. The lower temperature ordering is driven by ruthenium mediated interactions between holmium moments as spin ice correlations develop. The unsaturated order is compatible with a fragmented ferromagnetic state equivalent to pyrochlore kagome ice.
\end{abstract}

\maketitle

\begin{figure}[tp]
 \includegraphics[width = 1\linewidth]{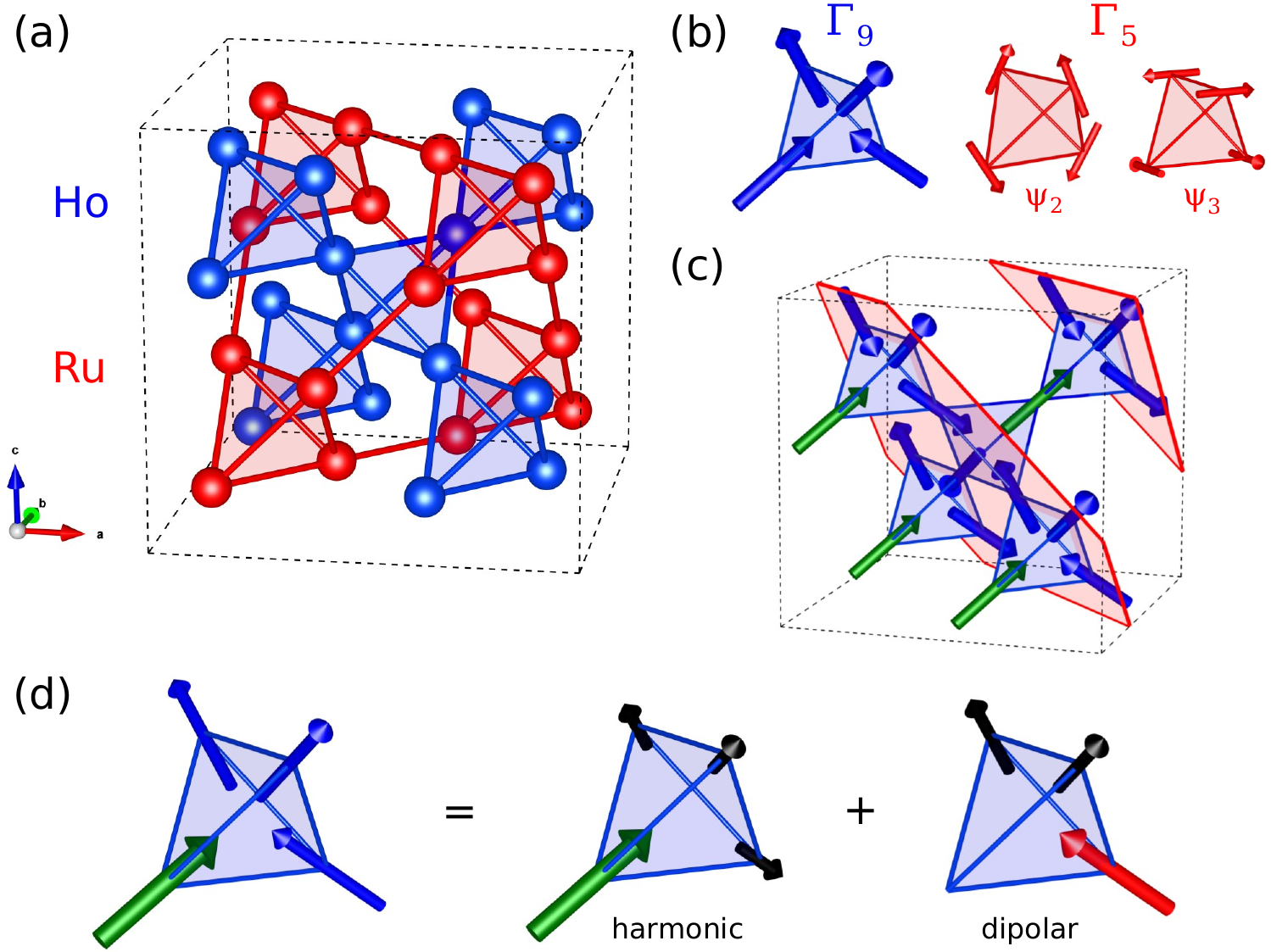}
 \caption{\label{Fig_structure} (a) The \HRO\ pyrochlore lattice; Ho sublattice in blue and Ru sublattice in red. (b) Ordered spin ice structure from the $\Gamma_9$ representation (left) and two states $\psi_2$ and $\psi_3$ from which easy-plane antiferromagnetic structures belonging to the $\Gamma_5$ representation can be built (right) (See for example Refs. \onlinecite{Wills2006, Poole2007, Petit2017} for the definition of these representations). (c) Ferromagnetic fragmented magnetic structure. (d) Graphical representation of the magnetic fragmentation (Equation \ref{eqFrag}). The green spin is the apical spin and the red spin is the majority moment of the dipolar fragment.}
\end{figure}

Highly frustrated magnets \cite{Gardner2010} have proved a rich source of materials showing novel, strongly correlated phases of matter. Among the most striking examples are holmium and dysprosium titanate, the original spin ice materials \cite{Harris1997,Bramwell2001} in which the magnetic Ho$^{3+}$ and Dy$^{3+}$ ions lie on a pyrochlore lattice, occupying the $A$ sites of the $A_2B_2$O$_7$ structure (Fig. \ref{Fig_structure}a). The $B$ sites, occupied by the non-magnetic Ti$^{4+}$ ions form a second, interpenetrating pyrochlore lattice.
Spin ices are frustrated ferromagnets in which strong crystal fields and combined exchange and dipole interactions result in effective first neighbour ferromagnetic exchange \cite{denHertog2000}.
The materials remain disordered down to 50 mK, way below the magnetic interaction scale, with the spins satisfying the ice rules of two spins pointing into and two pointing out of each tetrahedron \cite{Ramirez1999,Giblin2018}.  Great strides have been made in the understanding of these complex systems through the identification of an emergent gauge field description \cite{Isakov2004, Henley2005} with associated topological defects, magnetic monopoles \cite{Ryzhkin2005,Castelnovo2008}. This fractionalisation of magnetic moments into deconfined magnetic monopoles and the emergence of continuous symmetry from discrete, Ising like degrees of freedom, is captured in full by the process of fragmentation \cite{Brooks2014,Lhotel2020}. In this field theoretic formalism, the magnetic moments appear to separate, via a Helmholtz decomposition into a particle sector and left over gauge and harmonic sectors \cite{Bramwell2017,Museur2023}. 

Replacing the titanium by magnetic ions such as iridium or ruthenium adds to the richness of the emergent description.  The ions on the $B$ sites now order on the 100 K temperature scale \cite{Matsuhira2011, Ito2001}, providing strong internal fields that act on the $A$ sites at the energy scale associated with magnetic monopole physics. In the case of iridates, ``all in - all out'' ($\Gamma_3$) iridium ordering creates a strong staggered molecular field along the Dy$^{3+}$ or Ho$^{3+}$ moments which stabilises an ordered 3 out - 1 in / 3 in - 1 out state, although there is no further phase transition \cite{Lefrancois2017,Cathelin2020,Pearce2022}. Within the fragmentation picture the particle sector yields a monopole crystal with the zinc-blend structure, which manifests as Bragg peaks in neutron  experiments, and the gauge sector is that of emergent hard-core dimers on the diamond lattice of tetrahedra centres \cite{Huse2003,Brooks2014,Raban2019,Lhotel2020}, which results in a structured diffuse scattering. 

In ruthenates there is a further transition at low temperature (1.85 K and 1.55 K for Dy and Ho respectively) \cite{Ito2001,Bansal2002} but to an unsaturated phase again with a residual entropy \cite{Rams2011, Gardner2005}. Neutron diffraction experiments on a powder sample of \HRO\, were consistent with a ferromagnetic, $\Gamma_9$ spin ice state (See Fig. \ref{Fig_structure}b) with an ordered  moment estimated to be $6.3 \pm 0.5$~$\mu_{\rm B}$ per spin, considerably reduced from the full $10$~$\mu_{\rm B}$ per holmium ion \cite{Wiebe2004}. These two puzzling features have not been understood to date. In this paper, through neutron scattering, specific heat and magnetisation studies on a powder sample of  Ho$_2$Ru$_2$O$_7$, we show that this ordering is consistent with a novel, fragmented ground state with magnetisation aligned along one of the body centred cubic  $\langle111 \rangle$ axes as illustrated in Fig. \ref{Fig_structure}c. In addition, we observe in the {\it ac} susceptibility unusual dynamics, with a discontinuity in the characteristic relaxation times associated with the holmium transition.

\begin{figure}[tp]
    \includegraphics[width=1\linewidth]{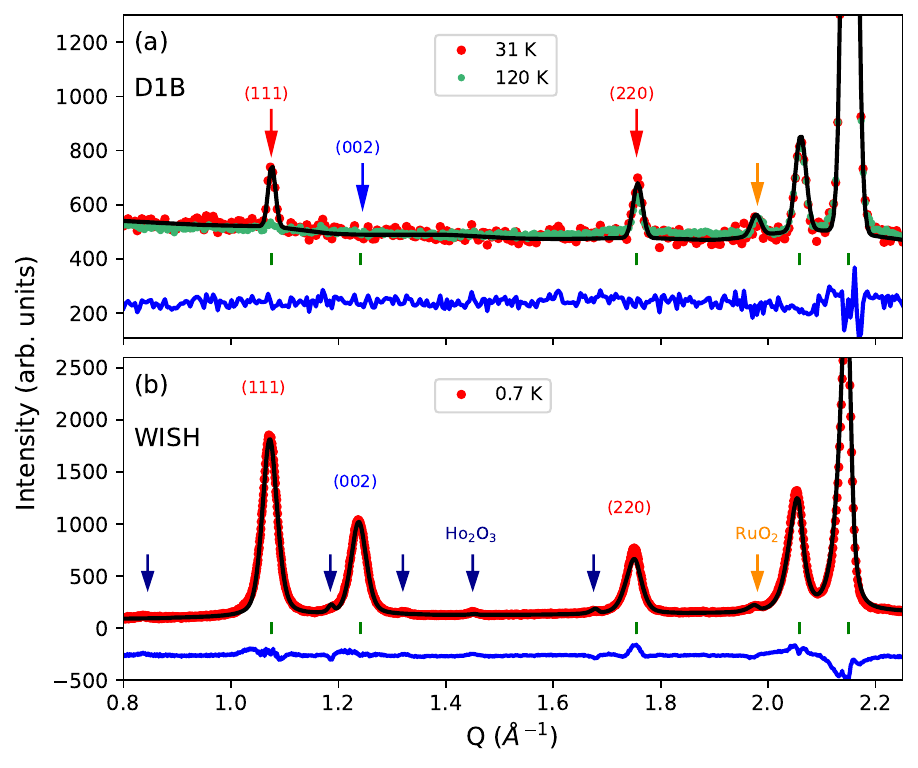}
    \caption{\label{Fig_diffraction} Neutron diffractograms: (a) at 31 and 120 K showing the Ru magnetic ordering; 
(b) at 0.7 K showing the Ho magnetic ordering. 
Black lines are the Rietveld refinements performed with the {\sc Fullprof} software \cite{FullProf1993}, with the $\Gamma_5$ and $\Gamma_9$ representations (See Fig. \ref{Fig_structure}) for Ru and Ho respectively \cite{supmat}. Blue lines are the differences between the fit and the data and green ticks index the Bragg peaks. The (002) peak, indicated by a blue arrow, is only present in the $\Gamma_9$ representation, and appears below the Ho transition. 
Dark blue and orange arrows indicate the peaks originating from the Ho$_2$O$_3$ and RuO$_2$ impurities respectively.}
\end{figure}

The sample was synthesized by solid state reaction from the oxide precursors Ho$_2$O$_3$ and RuO$_2$, with a small excess (5\% mol) of RuO$_2$ following Ref. \onlinecite{Gaultois2013} (See Supplementary Material \cite{supmat}). 
An X-ray diffraction analysis of the crystal structure post synthesis revealed the presence of Ho$_{2}$O$_{3}$ and RuO$_2$, visible as small peaks in neutron diffractograms (see Fig. \ref{Fig_diffraction}b).
Magnetometry measurements were performed on a MPMS VSM squid magnetometer between 2 and 300 K and SQUID magnetometers equipped with a dilution refrigerator developed at the Institut N\'eel between 70 mK and 4.2 K \cite{Paulsen2001}. Specific heat measurements were performed on a PPMS apparatus between 350 mK and 300 K. 
Neutron diffraction measurements were performed on D1B - CRG@ILL \cite{doi_D1B}, WISH@ISIS \cite{doi_WISH,Chapon2011} as well as D007@ILL with XYZ polarization analysis \cite{doi_D7,Nilsen2020}. % \cite{doi_D007}
Inelastic neutron scattering measurements were performed on LET@ISIS \cite{doi_LET1,doi_LET2,supmat}. 

Specific heat measurements show a transition at 95~K \cite{supmat} consistent with previous studies  \cite{Ito2001,Bansal2002}, but which is not apparent in magnetization measurements \cite{supmat}. In Fig.~\ref{Fig_diffraction}a we show neutron scattering data at 31 K which highlights the Ru ordering. The magnetic structure was previously interpreted as ferromagnetic, belonging to the  $\Gamma_9$ spin ice state \cite{Wiebe2004}. However, such an ordering would be characterised by a Bragg peak at the $(002)$ position \cite{Petit2017}, indicated by the blue arrow. 
Our data which profits from better statistics does not resolve this peak. In its absence, peaks at the $(111)$ and $(220)$ positions are consistent with an easy-plane $Q=0$ state belonging to the $\Gamma_5$ representation (See Fig. \ref{Fig_structure}b), as reported in other pyrochlore ruthenates \cite{Ito2001,Taira2003}.
Note that the exact spin configuration within the $\Gamma_5$ manifold cannot be obtained from powder neutron diffraction \cite{Poole2007}.
The intensity of the magnetic peaks saturates below $75$~K and a Rietveld refinement gives an ordered ruthenium moment of around $1.2 \pm 0.2~\mu_{\rm B}$. 

The holmium magnetic structure at low temperature (See Fig. \ref{Fig_diffraction}b), is consistent with ferromagnetic spin ice belonging to the $\Gamma_9$ representation,  with a refined ordered moment of around $6.5\pm 0.4~\mu_{\rm B}$ per Ho at 300 mK \cite{supmat}, similar to Ref \onlinecite{Wiebe2004}. 
Down to the base temperature, the magnetic Bragg peaks remain broader and more symmetric than the nuclear peaks showing the characteristic time-of-flight lineshape. This indicates that the order is made up of mesoscopic domains, which were estimated from the full width at half maximum of the $(002)$ peak to be of the order of 100 \AA, corresponding to roughly ten elementary cubic units or 16000 spins.

\begin{figure}[tp]
    \centering
    \includegraphics[width=1\linewidth]{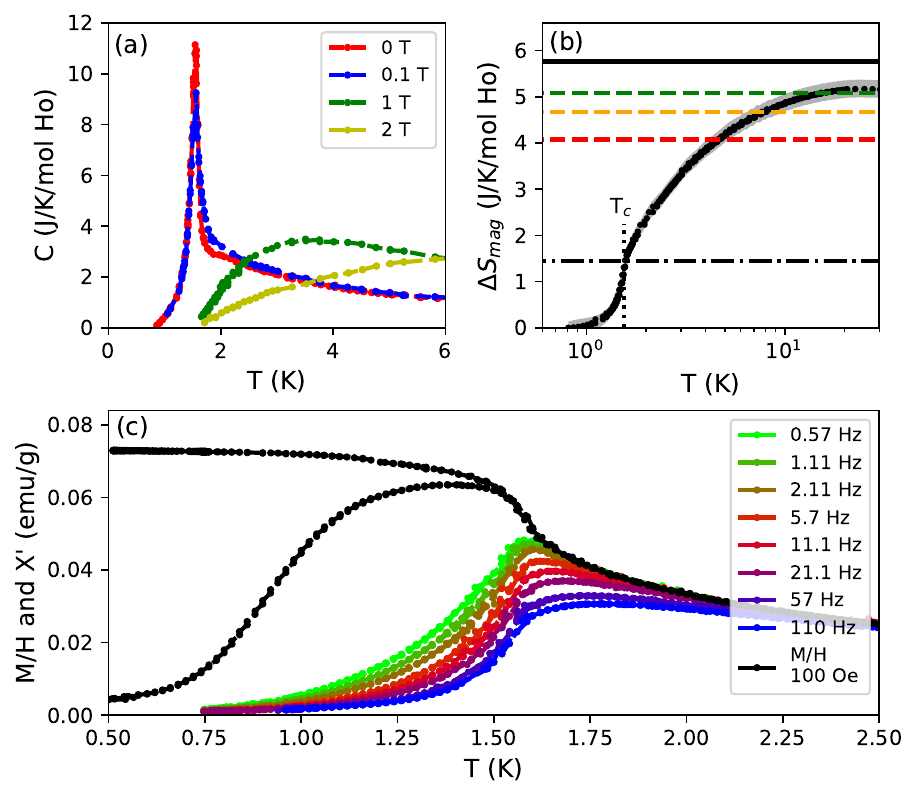}
    \caption{\label{Fig_CM_LT}(a) Magnetic specific heat in zero and applied magnetic fields, corrected for nuclear hyperfine contribution and for the phonon signal using a non magnetic sister compound \cite{supmat}. (b) Magnetic entropy difference with the expected values for spin ice (red), monopole crystal (orange), kagome ice (green), $R\ln(2)$ (full black), $\frac{1}{4}R\ln(2)$ (dashed black). (c) In-phase part of the {\it ac} susceptibility $\chi'$ as a function of temperature, together with the magnetization to field ratio measured in a ZFC-FC protocol with a field of 100 Oe (black). }
\end{figure}

The sample enters the Ho ordered phase via a transition at $T_c \approx 1.55$~K with a steep rise in the magnetic Bragg peaks intensity, a marked increase in the magnetization, and a very sharp peak in the magnetic specific heat  (See Fig. \ref{Fig_CM_LT}a and c). This temperature is the typical interaction scale for spin ice physics in \HTO\ \cite{Bramwell2001}.
Integrating over the specific heat peak we find a residual entropy at low temperature (See Fig. \ref{Fig_CM_LT}b), as previously reported \cite{Gardner2005}. 
The singularity resists fields up to $0.1$ T but is smoothed out for higher fields.

\begin{figure}[tp]
    \centering
    \includegraphics[width=1\linewidth]{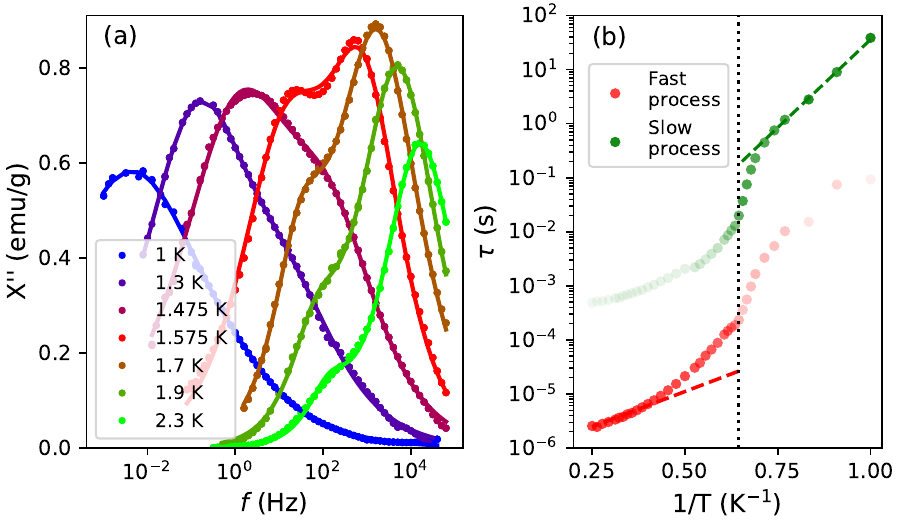}
    \caption{\label{fig_Xac} (a) $\chi"$ as a function of frequency at fixed temperatures. (b) Relaxation times $\tau$ from a fit to a generalized Debye function with two characteristic times \cite{supmat}. The transparency of the symbols indicate the relative amplitude of the two processes. Dotted lines are Arrhenius laws $\tau_0 \exp(E/T)$, with $\tau_{0f}=5.5 \times 10^{-7}$ s and $E_f=6$ K above $T_c$ and $\tau_{0s}=1.1 \times 10^{-5}$~s and $E_s=15$ K below.}
\end{figure}

A large Zero Field Cooled - Field Cooled (ZFC-FC) irreversibility develops in the magnetization below the transition, indicating the presence of significant energy barriers. {\it Ac} susceptibility measurements as a function of temperature show the presence of  frequency dependent dynamic behaviour, which persists below the transition. 
At fixed temperature, the imaginary part of the {\it ac} susceptibility, $\chi"(f)$ (See Fig. \ref{fig_Xac}), is characterized by two peaks, present over the whole temperature range, corresponding to a rapid and a slow process. This signal is reminiscent of that for \HTO\ \cite{Wang2021,Morineau2024}, and the persistence of well defined dynamic behaviour below the ordering temperature calls to mind  experiments on \HIO\ \cite{Lefrancois2017}. The two time scales can be extracted from our susceptibility data from a generalized Debye analysis \cite{supmat} (see Fig. \ref{fig_Xac}b). The relative amplitudes of the two processes vary with temperature. Above the transition, the dominant fast process approaches an Arhenius law with an energy of about 6 K, close to the value of the monopole chemical potential in \HTO, $-\mu=5.8$~K \cite{Castelnovo2008,Jaubert2011}. This suggests relaxation via the dynamics of deconfined monopoles. At the transition the amplitudes are equal and the transition signifies a change in the dominant relaxation process. Below the transition the prevailing slow process evolves towards a characteristic energy scale of 15~K. These persistent and sizeable dynamics at low temperature point to an unconventional ordered state with well-defined excitations.

The sharp specific heat peak, domains of modest size, rounded step function for the ordered moment and changing dynamics through the transition are reminiscent of a rounded first order transition, the rapid fall in entropy through the transition being the vestige of a latent heat.
The transition could still be second order but with an extremely narrow critical region. It is clearly correlated with the suppression of monopoles, so that the resulting topological constraints could lead to an unconventional topological transition \cite{Jaubert2010}.

The holmium ferromagnetic transition is driven by ruthenium mediated interactions between holmium moments. In the $\Gamma_5$ structure, the Ru$^{4+}$ ions create molecular fields at the Ho$^{3+}$ sites lying in the local $(x,y)$ planes so that there is no coupling to the component along the $z$-axis. In fact, as Ho$^{3+}$ is a non-Kramers ion it is totally blind to these fields. It is this null result that allows the system to have a second symmetry breaking phase transition. Ferromagnetic interactions are induced through the relaxation of the ruthenium moments out of their easy planes, driven by the build up of holmium-holmium correlations. This is quite different from the iridates \cite{Lefrancois2017,Cathelin2020,Pearce2022} where the molecular fields from the ordered iridium moments break symmetry for the $A$ site moment, in favour of a $\Gamma_3$ state at all temperatures resulting in a magnetic crossover only. We have modelled this using a nearest neighbour Hamiltonian with a spin ice term between the Ho moments, a local easy-plane coupling between the Ru ions and a ferromagnetic coupling between local $z$ components of neighbouring Ho and Ru ions \cite{supmat}. We find that, as Ho spin ice correlations grow the Ru moments indeed tilt out of their easy planes, satisfying the ice rules on each $B$ tetrahedron and generating a fully ordered $\Gamma_9$ state for the Ho moments.

As this mechanism leads to a saturated state it cannot be the full story.  A first possibility is that the system  forms a mosaic of ordered domains with wide, disordered domain walls accounting for the missing moment and the configurational entropy. In the absence of magnetic monopoles and with the imposition of topological constraints, domain wall boundaries will play a very influential role \cite{Bovo2014,Jaubert2017,Lantagne2018},
with ``arctic-circle'' freezing effects running deep into the bulk \cite{Allegra2016, King2023}. The origin of the domains could be dynamical, following the freezing out of monopoles, or it could be static due to the formation of Weiss domains driven by the long range part of the dipolar interactions.

A second possibility is that the system orders into a different ferromagnetic state with unsaturated order parameter and finite entropy. We have remarked that the residual entropy corresponds to that for the magnetisation plateau for spin ice in the presence of a field applied along the $[111]$ crystal axis \cite{Matsuhira2002, Moessner2003}, the so-called kagome ice state as shown in Fig. \ref{Fig_CM_LT}b where we also show the Pauling entropy \cite{Ramirez1999} and that for a fragmented monopole crystal \cite{Brooks2014}. 
The residual entropy is therefore consistent with a ferromagnetic phase with broken symmetry along one of the $\langle 111 \rangle$ axes, {\it the ferromagnetic fragmented state} which features decoupled kagome planes of magnetic moments lying perpendicular to the ordering direction. 
The dynamics, with well defined time scales which persist below the transition also seems consistent with such a structured state, rather than a random mosaic state where one would expect a broader range of time scales and  glassy behaviour at low temperature.

This phase requires a distinction between the four spins of a unit cell, as highlighted in Fig. \ref{Fig_structure}c. The apical spin (shown in green) would be frozen and aligned along the broken symmetry axis with an entropy drop of approximately $\left(\frac{R}{4}\right)\ln 2$, as observed in our data at the transition. The three spins lying in the kagome plane (in red) would then satisfy the kagome ice rules of two spins in - one spin out \cite{Moller2009,Chern2011}. 

This is another example of a fragmented ground state in which identifiable ordered and disordered fragments coexist. Here, the ground state is a monopole vacuum but the spins still fragment into two parts: a gauge, or dipolar sector \cite{Brooks2014,Lhotel2020} contributing closed internal loops of magnetic flux, and a harmonic sector \cite{Museur2023} providing the (unsaturated) ferromagnetic moment (See Fig. \ref{Fig_structure}d). 
The four holmium moments can be thought of as needles joining tetrahedron centre $I$ to neighbouring centres $J$ \cite{Castelnovo2008,Brooks2014}, carrying magnetic flux $M_{IJ}$ and
can be decomposed in this spirit. 
Taking the first element to represent the apical spin:
\begin{align}\label{eqFrag}
\left[M_{IJ}\right]&= \quad    \left[ 1, 1, -1, -1 \right] \nonumber \\
    &=\quad \left[1, -\frac{1}{3}, -\frac{1}{3}, -\frac{1}{3} \right] &+& \quad \left[0, \frac{4}{3}, -\frac{2}{3}, -\frac{2}{3}\right]\\
    &=\qquad \quad  [M_{IJ}]_h&+&\qquad \quad [M_{IJ}]_d \nonumber,
\end{align}
where $M_{IJ}=-M_{JI}$ and where an element entering a tetrahedron is negative. 
The configurations are divergence free. The dipolar fragments, $[M_{IJ}]_d$ whose loops give the residual entropy are restricted to the kagome planes \cite{Moessner2003} and the harmonic fragments, $[M_{IJ}]_h$ are identical for each unit cell. 

\begin{figure}[tp]
    \centering
    \includegraphics[width=1\linewidth]{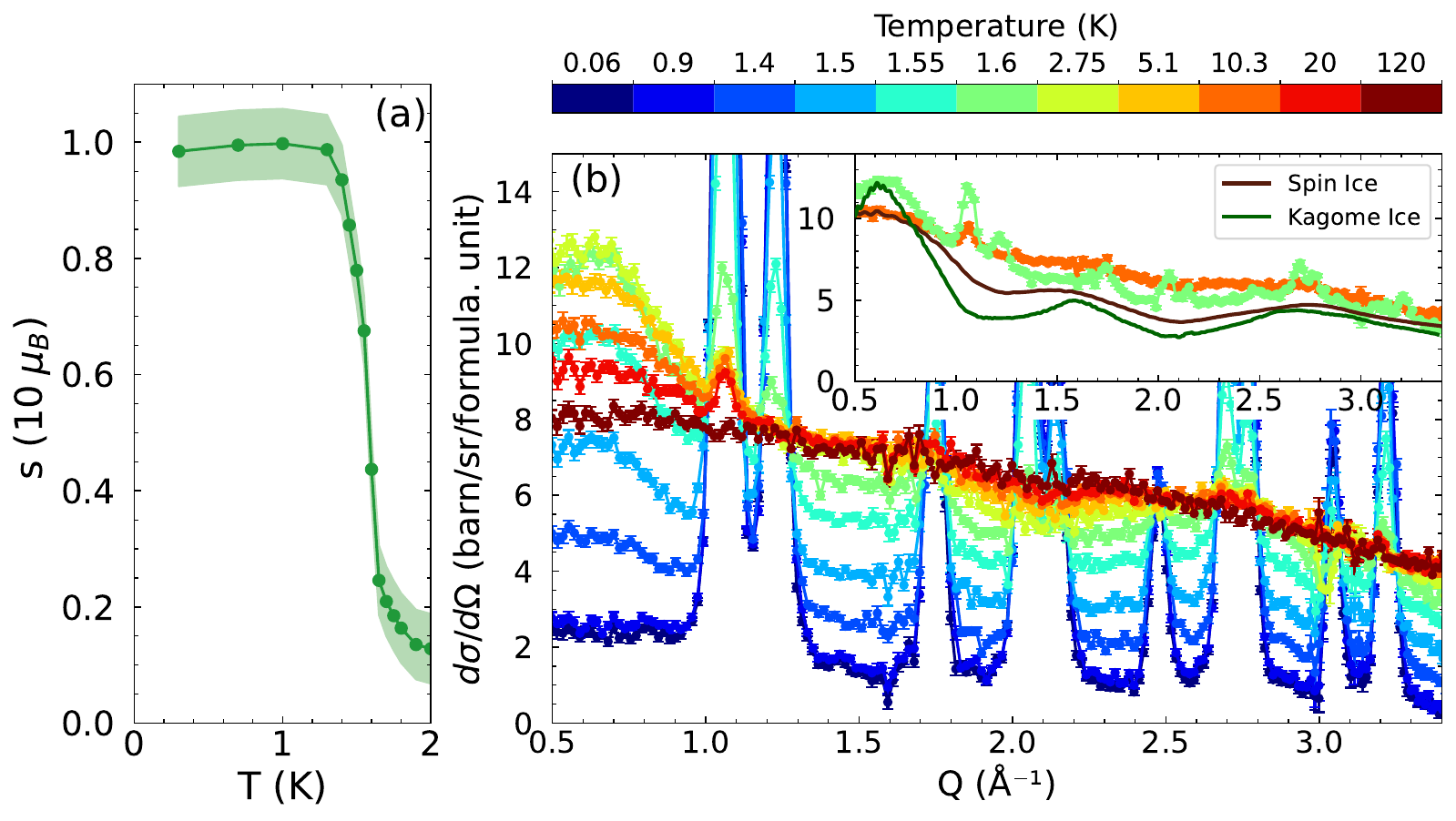}
    \caption{ \label{fig_mord} (a) Refined component $s$ as a function of temperature in the proposed fragmented structure (See text). The shaded region shows the error bars. (b) Magnetic neutron scattering intensity measured on D007@ILL between 60 mK and 120 K. Inset: Zoom in the magnetic diffuse neutron scattering above the transition temperature at 1.6 and 10 K, compared to the pattern expected for the kagome ice and spin ice scattering functions.}
\end{figure}

This state is a combination of three partially ordered $\Gamma_9$ states aligned along orthogonal $\langle 100 \rangle$ axes \cite{Museur2023} so that in a powder sample the magnetic Bragg peaks are identical to those of a $\Gamma_9$ state. As a consequence the two possibilities cannot be distinguished directly in our experiment, although the refinement procedures are quite distinct and the estimate for the ordered moment depends on which is used. 
We have refined our diffractograms assuming the fragmented state, considering a thermal spin length $s$: 
\begin{equation}
[M_{IJ}]_h=\left[s,-\frac{s}{3},-\frac{s}{3},-\frac{s}{3}\right].
\label{order-m}
\end{equation}
The moment on the apical spin is then $s \times m_{\rm Ho} \approx s \times10~ \mu_{\rm B}$. As shown in Fig. \ref{fig_mord}a, we observe a sharp rise in $s$ through the transition to a value close to $s=1$, leaving the moment on the apical spin saturated and an average 
 moment per spin of approximately 5 $\mu_{\rm B}$.
This refinement, leading to $s=1$ at low temperature therefore provides quite a stringent test in favour of ordering into the fragmented ferromagnetic state.

The selection of this state would need an additional term in the model Hamiltonian which pushes the magnetisation away from the $\langle 100 \rangle$ axes towards the body diagonals. At the molecular field level, this corresponds to a repulsive self-consistent field with cubic symmetry \cite{supmat}. If the relaxation of the Ru$^{4+}$ moments out of their $\Gamma_5$ planes is constrained to follow the three in - one out ice rules of a monopole crystal \cite{Brooks2014} then such a cubic term would emerge. Such a constraint could originate in the interaction between the magnetic and electronic degrees of freedom on the Ru sublattice as proposed for iridate compounds \cite{Goswami2017, Ladovrechis2021}. 

A key feature of the fragmented ferromagnet is that the dipolar fragment should maintain a diffuse scattering signal \cite{Brooks2014}, characteristic of kagome ice \cite{Tabata2006, Moessner2003}. 
In Fig. \ref{fig_mord}b we show our diffuse scattering data for varying temperatures. Two trends are observed. Below about 20 K, magnetic diffuse scattering develops, and is characterised by a broad peak compatible with a powder of spin ice down to 2.75 K \cite{Hallas2012}.  On lowering the temperature further, close to the transition, the peak sharpens around $Q=0.57$ \AA$^{-1}$, a value characteristic of kagome ice \cite{Moessner2003}. Below the transition, the trend is reversed and the intensity falls.  The peak at $Q=0.57$ \AA$^{-1}$ remains visible down to 1.4 K before the spectrum flattens out at the lowest temperatures. The remnant liquid-like diffuse scattering at all temperatures below the transition is as expected for a phase with an unsaturated ordered moment. There is therefore some evidence of the build up of kagome ice correlations through the transition but the weak intensity at low temperature excludes a quantitative conclusion.

In conclusion, we have shown that in \HRO\ the Ru$^{4+}$ ions order in the $\Gamma_5$ easy plane antiferromagnetic state at around 100 K. This in turn drives a transition for the Ho$^{3+}$ ions into a partially ordered ferromagnetic state at $1.55$ K. The mechanism for the second transition is indirect, requiring collectively organised tilting of the Ru$^{4+}$ ions out of their easy plane. The partial order could be due to the formation of a mosaic of domains ordered along the $\langle100\rangle$ directions, frustrated by domain boundaries and non-ergodicity. However, the scale of the 
ordered moment, the residual entropy and the well-defined dynamics from {\it ac} susceptibility measurements are rather compatible with a specific fragmented ferromagnetic state with the moment lying along the $[111]$ direction and we offer a phenomenological ordering mechanism that could drive such a state. 
We hope that the present work will motivate future projects to address these questions. In particular, the recent synthesis of single crystals \cite{Patel2023} is promising, while understanding the ordering mechanism requires insight beyond lowest order perturbation theory. 

%%%%%%%%%%%%%%%%%%%%%%%%%%%%%%%%%%%%%%%%
% Acknowledgment
%%%%%%%%%%%%%%%%%%%%%%%%%%%%%%%%%%%%%%%%
\begin{acknowledgments}
We thank Michel Gingras, Michel Kenzelmann, Sylvain Petit, Jean-Marie St\'ephan and Slava Rychkov for useful discussions. We thank Pierre Lachkar for technical support on the PPMS and Carley Paulsen for the use of his magnetometers. F. Morineau acknowledges financial support from the LANEF Ph.D. Program. N. B. acknowledges financial support of the Program QuanTEdu-France n$^{\circ}$ ANR-22-CMAS-0001 France 2030. We acknowledge financial support from the ``Agence Nationale de la Recherche'' under Grant No. ANR-19- CE30-0040 and partially under grant NSF PHY-2309135 to the Kavli Institute for Theoretical Physics (KITP). PCWH also thanks the IHES for hospitality. 
\end{acknowledgments}

%%%%%%%%%%%%%%%%%%%%%%%%%%%%%%%%%%%%%%%%
%Bibliography
%%%%%%%%%%%%%%%%%%%%%%%%%%%%%%%%%%%%%%%%
\bibliography{biblio-arxiv}

%%%%%%%%%% Merge with supplemental materials %%%%%%%%%%
\onecolumngrid%\widetext
\clearpage

\begin{center}
\textbf{Ferromagnetic fragmented state in the pyrochlore \HRO : \\
Supplemental Material}
\end{center}

%%%%%%%%%% Prefix a "S" to all equations, figures, tables and reset the counter %%%%%%%%%%

\setcounter{section}{0}
\renewcommand{\thesection}{S-\Roman{section}}

\setcounter{equation}{0}
\renewcommand{\theequation}{S\arabic{equation}}

\setcounter{figure}{0}
\renewcommand{\thefigure}{S\arabic{figure}}
\renewcommand{\theHfigure}{S\arabic{figure}}

\setcounter{table}{0}
\setcounter{page}{1}

\makeatletter

\section{Synthesis, structural properties and diffraction data refinement}
The sample was synthesized by solid state reaction from the oxide precursors Ho$_2$O$_3$ and RuO$_2$, with a small excess (5\% mol) of RuO$_2$ following Ref. \onlinecite{Gaultois2013}. The powders were finely ground, pressed in a pellet and reacted at 1000 $^{\circ}$C in air for 48 h, then 1100$^{\circ}$C in a platinum tube inside an evacuated sealed silica tube for 12 days with intermediate grindings. 

X-ray measurements at ambient temperature show that the \HRO\ crystallizes as expected in the Fd$\bar{3}$m pyrochlore structure, with a lattice parameter $a=10.14$ \AA. Impurity phases of Ho$_2$O$_3$ (Ia$\bar{3}$, $a=10.61$ \AA) and RuO$_2$ (P4$_2$/mnm, $a=4.49$ \AA, $c=3.10$ \AA) were detected in our sample. The final composition is: (i) \HRO: 97.3(3) \% (ii) Ho$_2$O$_3$: 1.39(4) \% (iii) RuO$_2$: 1.27(4) \%.

\begin{figure*}[htb]
        \centering
        \includegraphics[width=0.65\linewidth]{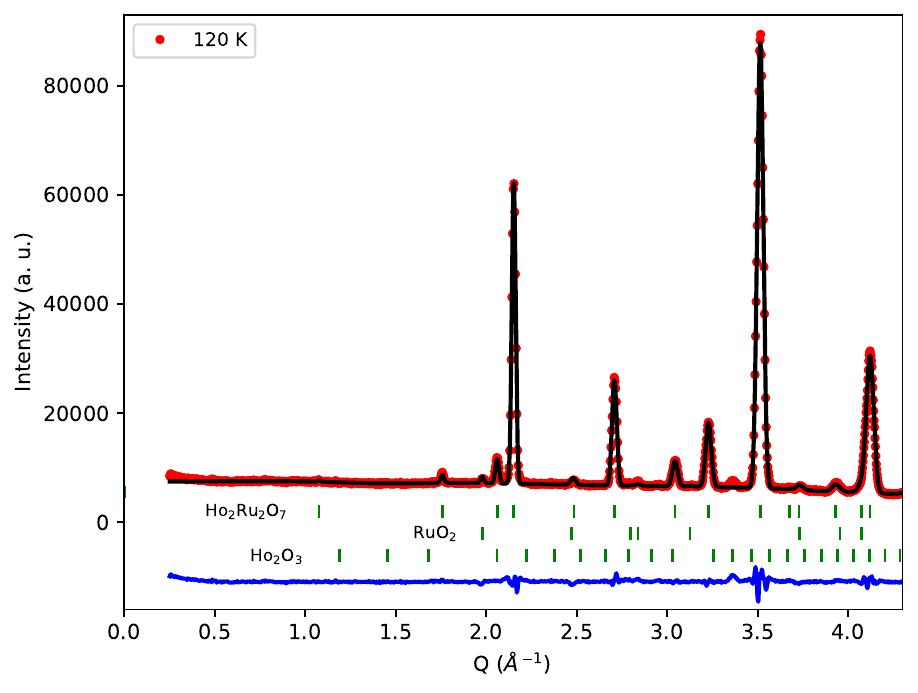}
       \caption{Diffractogram measured at 120 K on D1B,  together with the structural refinement including the two impurity phases of Ho$_2$O$_3$ and RuO$_2$.}
        \label{fig:diffracto}
    \end{figure*}

Neutron diffraction measurements were performed on the diffractometer D1B@ILL, with $\lambda=2.52$ \AA\ on a 3 g powder sample. The sample was contained in a vanadium sample holder and cooled down with a $^4$He orange cryostat. 
The diffractogram was first refined at 120 K, above the temperature of the magnetic transitions, using the {\sc Fullprof} software and the expected Fd$\bar{3}$m pyrochlore structure was recovered (See Figure \ref{fig:diffracto}). At 10 K, the parameter for the position of the oxygen is $x=0.335$ and the lattice constant is $a=10.11$ \AA.

Similar parameters were later obtained on the time-of-flight diffractometer WISH@ISIS. A magnetic signal from Ho$_2$O$_3$ could be observed below $1.2$ K in these measurements.
\medskip

The neutron diffraction data refinement, using the FullProf software, was carried out as follows:
\begin{itemize}
    \item Above 100 K, the position of the zero angle as well as the scales of the sample phase and sample holder phases were defined.
    \item Between 10 and 100 K, the Ru magnetic structure was refined together with the lattice parameters. Both $\psi_2$ and $\psi_3$ basis states from the $\Gamma_5$ representation were tested and yielded the same result. Below about 70 K, the Ru moment stayed constant at $1.2~\mu_{\rm B}$ per atom.
    \item Below 10 K, both Ho and Ru magnetic structures were refined in the same phase, as to take into account possible interference effects. Leaving the Ru moment as a variable did not significantly improve the quality of the fit, so it was left fixed at its 10~K value. A Lorentzian peak width parameter was added to take into account the broadening observed at low temperature. The lattice constant was kept fixed from the values determined at 10~K. The Ho magnetic structure was refined as a harmonic fragment defined in Equation 2 of the main text. To improve the fit, we tried adding various magnetic tilts of allowed symmetries to the Ru magnetic structure, but this didn't show any improvements. It was therefore kept fixed at its 10 K value.
\end{itemize}

\begin{figure*}[htb]
        \centering
        \includegraphics[width=0.65\linewidth]{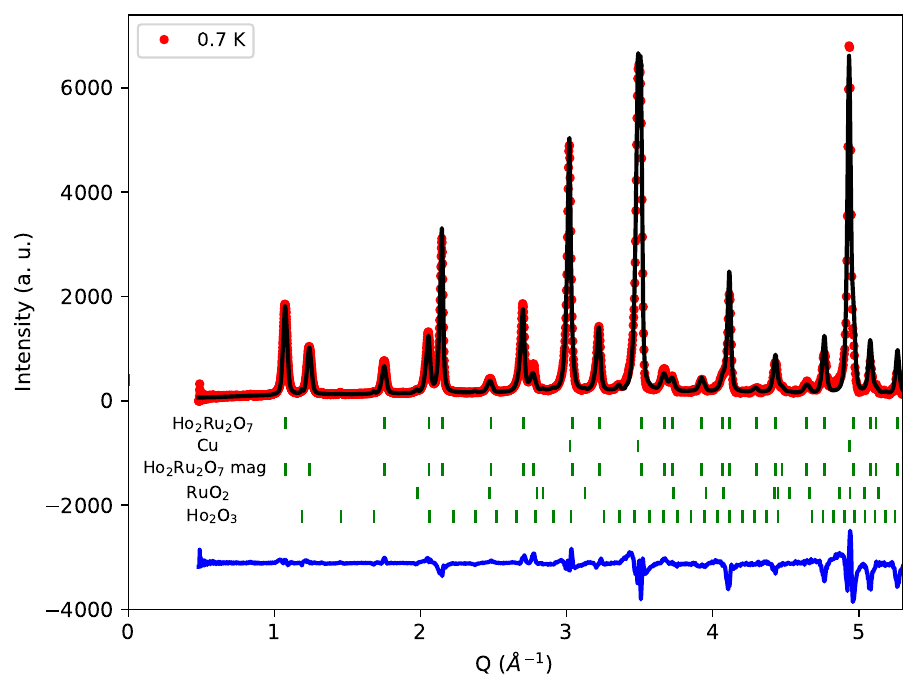}
       \caption{Diffractogram measured at 0.7 K on WISH, together with the structural and magnetic refinements. The magnetic phase is refined with a fragmented ferromagnet phase as detailed in the main text. A phase for the copper from the calorimeter is included as well as Ho$_2$O$_3$ and RuO$_2$ impurity phases.}
       \label{fig:0p7K}
    \end{figure*}

In the WISH data, an absorption parameter was included in the fits. It allows to take into account the effect of the absorption in the time-of-flight (TOF) experiments which is stronger at low $Q$ (large TOF) than at high $Q$ (small TOF). 
The resulting fit obtained for $T=0.7$ K is shown on figure \ref{fig:0p7K}.

Polarized neutron scattering diffraction measurements were performed on D007@ILL with $\lambda=2.95$ \AA\ on a 1.9 g powder sample in a copper annular sample holder, cooled down to 60 mK in a dilution fridge. XYZ polarization analysis were performed using the 6-point method, after corrections using vanadium, quartz and background references \cite{Stewart09}. Before this experiment, new thermal treatments were performed on the sample, which allowed to reach a lower impurity level: 1.17(3) \% for Ho$_2$O$_3$ and 0.36(2) \% for RuO$_2$. 

\section{``High" temperature characterization}
\begin{figure*}[htb]
        \centering
        \includegraphics[width=0.329\linewidth]{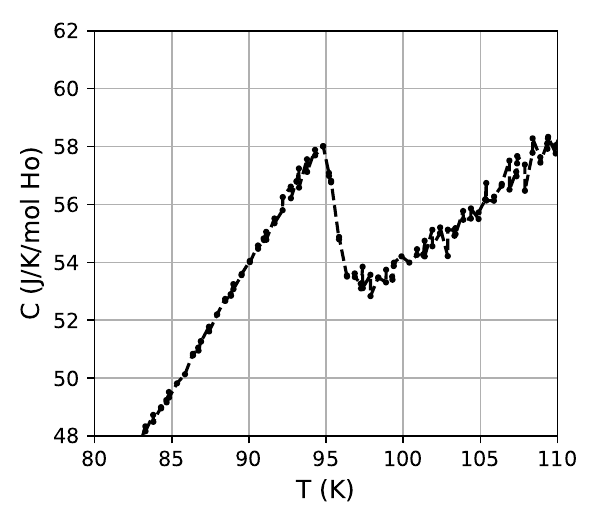}
        \includegraphics[width=0.329\linewidth]{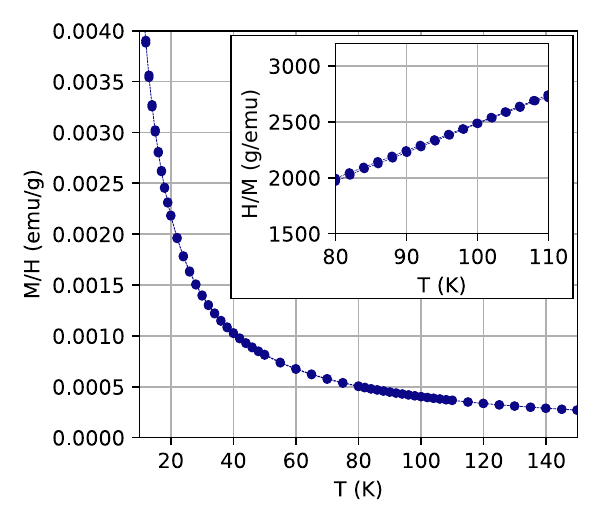}
        \includegraphics[width=0.329\linewidth]{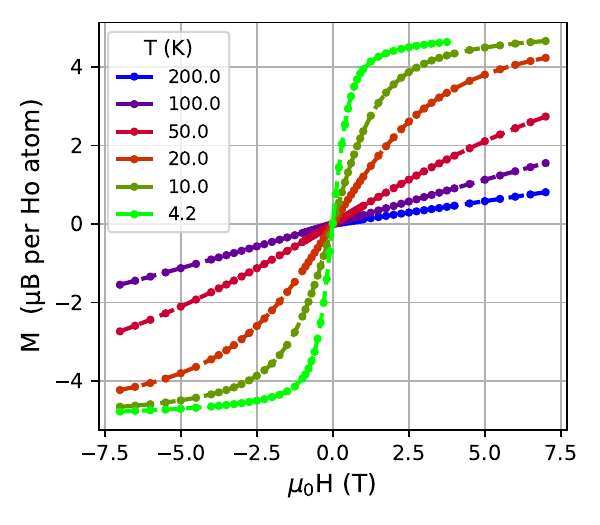}
        \caption{(left) Specific heat $C$ vs temperature $T$ showing the Ru transition at 95 K. (middle) $M/H$ measured at  1000 Oe with a ZFC-FC protocol. The inset shows $H/M$ around the Ru transition for the same dataset. (right) Magnetization $M$ vs field $H$ for several temperatures between 4.2 and 300 K.}
        \label{fig:HRO_HT}
    \end{figure*}

On Figure \ref{fig:HRO_HT} we show magnetic and calorimetric characterization measurements of our \HRO\ sample above 2~K, performed on PPMS and MPMS setups respectively. The specific heat (left) shows an anomaly at about $95$~K which corresponds to the temperature of the ruthenium transition. The magnetization does not show any change of curvature or ZFC-FC effect around the ruthenium transition at 95~K, contrary to what was seen in Ref. \onlinecite{Bansal2002} (middle figure). The transition is not visible either in a $\mathrm{d}M/\mathrm{d}T$ analysis. The signature of the Ru transition is expected to be small in magnetization measurements, due to its small moment compared to Ho. In the case of a ferromagnetic transition, a small step would nevertheless be awaited in the magnetization. Its absence is thus consistent with the antiferromagnetic $\Gamma_5$ Ru ordering determined in neutron diffraction experiments. 

In pyrochlore iridates, the ZFC-FC effect at the Ir transition has been attributed to the uncompensated moments at the boundaries between domains \cite{Stasko2024} and is possibly reinforced when pinning is present due to defects \cite{Zhu2014, Lefrancois2015}. The absence of such an effect in our \HRO\ sample could be the sign of a good quality of the sample, regarding the local environments of the magnetic ions.

\section{``Low" temperature magnetic properties}
  \begin{figure*}[htb]
        \centering
        \includegraphics[width=0.329\linewidth]{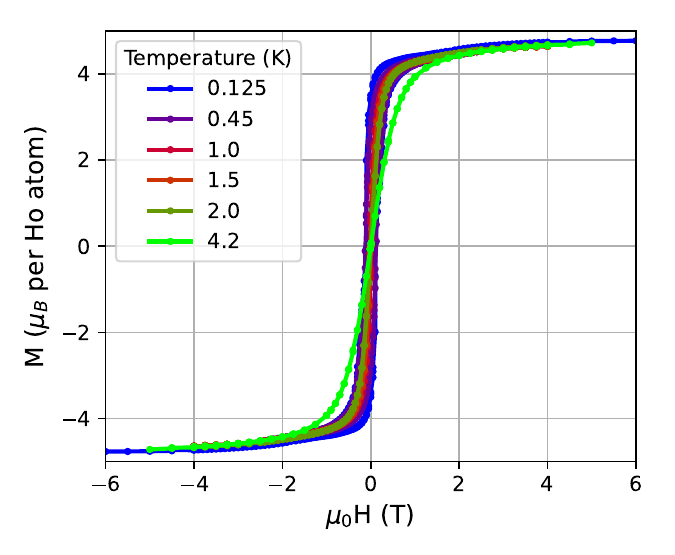}
        \includegraphics[width=0.329\linewidth]{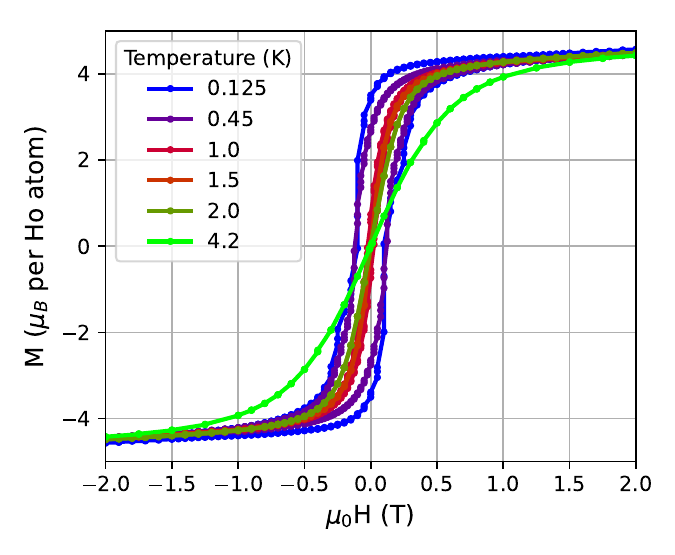}
        \includegraphics[width=0.329\linewidth]{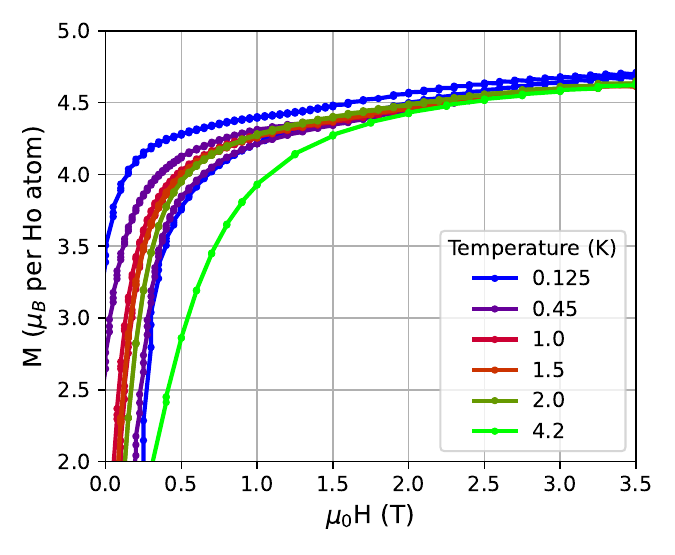}
          \caption{$M$ vs $H$ at several temperatures between 125 mK and 4.2 K: (left) up to 6 T; (middle) zoom in the hysteresis; (right) zoom in the inflection point. }
         \label{fig:HRO_MH_LT}
    \end{figure*}

 At 125 mK and 6 T, the magnetization reaches about 4.75~$\mu_{\rm B}$ per Ho atom (See Figure \ref{fig:HRO_MH_LT}(left)). For a pyrochlore magnet with a strong Ising anisotropy along the local $\langle 111 \rangle$ axes, the powder average saturation magnetization per spin is one half of the magnetic moment $\mu_{\parallel}$ \cite{Bramwell2001b}. Our measurement is thus consistent with a Ho$^{3+}$ ion with a strong Ising anisotropy. The corresponding Ho magnetic moment $\mu_{\parallel}$ would be 9.5~$\mu_{\rm B}$. This value is small compared to the 9.8-9.9~$\mu_{\rm B}$ value reported in other Ho pyrochlores and obtained from the crystal electric field calculations. This difference may be due to the presence of the impurity phases, whose mass has not been taken into account.
   
       \begin{figure*}[h!]
        \centering
        \includegraphics[width=0.4\linewidth]{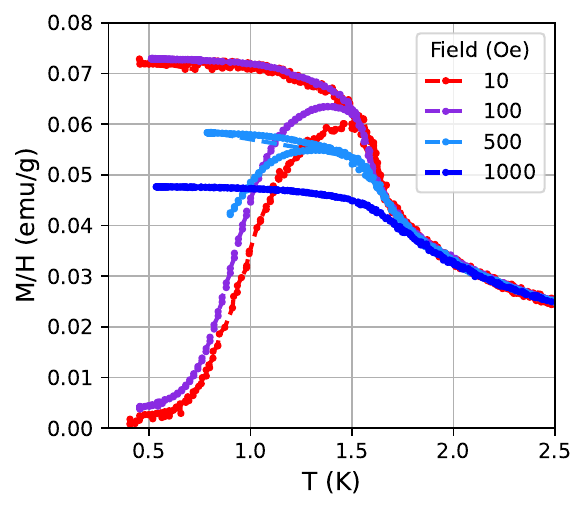}
        \caption{$M/H$ vs $T$ measured in a ZFC-FC protocol at several fields. The ZFC state was cooled down from 4.2 K. }       
         \label{fig:HRO_MT_LT}
    \end{figure*}

An irreversibility is observed at low temperature between the ZFC and FC curves, where the field-cooled behavior suggests a ferromagnetic transition (See Figure \ref{fig:HRO_MT_LT}). Given the scale of the effective ferromagnetic exchange in Ho pyrochlores ($J_{\text{eff}} \approx 2$~K), it is natural to associate this transition to a magnetic ordering on the Ho sublattice. The amplitude of the ZFC-FC effect is large, indicating the presence of significant energy barriers below the transition. The irreversibility is suppressed when increasing the magnetic field and has disappeared for $H=3000$ Oe.

An hysteresis is observed in the $M$ vs $H$ curves below the Ho transition (See Figure \ref{fig:HRO_MH_LT}(middle)). An anomaly is observed around 1.7 T close to the saturation at low temperature (See Figure \ref{fig:HRO_MH_LT}(right)). The interpretation of this effect is hard due to the powder average, but it could be reminiscent from the transition observed between the kagome ice state and the saturated monopole crystal when a field is applied along $[111]$, in spin ices \cite{Sakakibara2003, Petrenko2003} and Ho$_2$Ir$_2$O$_7$ \cite{Lefrancois2017}.

\section{Specific heat data analysis}
\begin{figure}[htb]
        \includegraphics[width=0.4\linewidth]{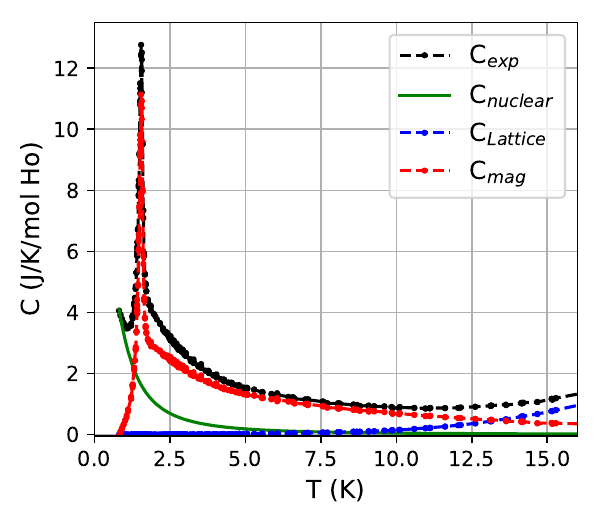}
        \caption{Measured specific heat in black, nuclear contribution in green, lattice contribution in blue, and subtraction in red.}
    \label{fig:Ccontrib}
\end{figure}

The magnetic specific heat of \HRO\ is shown in the main article, Figure 3. 
To obtain this magnetic contribution in the temperature range $< 25$~K from the total specific heat (see Figure \ref{fig:Ccontrib}), one must subtract the contributions from both the nucleus and the lattice. 

Below $1$ K, the rise in specific heat is caused by the hyperfine contribution coming from the $^{165}$Ho nucleus. Its $I= 7/2$ nuclear spin is split by the interaction with the electronic spin. The corresponding contribution can be obtained from a $2I+1 =8$ level system obeying the following Hamiltonian:
\begin{equation}
    \mathcal{H}_N = A_{\parallel} I_z + P  \left( I_z^2 - \frac{1}{3} I(I+1) \right)
\end{equation}
$A_{\parallel}$ is the strength of the hyperfine interaction, which splits the nuclear levels evenly, and $P$ is the quadrupole interaction, driven by the average quadrupole electronic moment, which modifies the splitting between high and low $I_z$. The computation of the associated specific heat for one nucleus is straightforward:
\begin{equation}
    C_N = \frac{\partial \langle E \rangle }{\partial T} = \frac{\partial }{\partial T} \left( k_B T^2 \frac{\partial \ln Z}{\partial T} \right) = \frac{\sum_{i=-I}^{I}\sum_{j=-I}^{I}(E_{i}^2-E_{i}E_{j}) \exp (\frac{-E_{i} - E_{j}}{k_B T}) }{ (k_B T) ^2 \sum_{i=-I}^{I}\sum_{j=-I}^{I}\exp (\frac{-E_{i} - E_{j}}{k_B T})}
\end{equation}
This expression could in theory be fitted to the data, but we could not accurately collect enough data points at low temperature to obtain a reliable fit. Indeed, below approximately 700~mK, the increase in specific heat and the decrease in grease conductivity cause the sample to become decoupled from the sample holder. This effect is amplified by the fact that we have a powder sample.

For this reason, we chose to manually adjust the rise in specific heat measured below 1~K, comparing our estimates of  $A_{\parallel}$ and $P$ with published values measured in other Ho$^{3+}$ systems \cite{Lounasmaa1962,Bramwell2001b,Mennenga1984}
: $A_{\parallel} \in \left[0.3, 0.4\right], \, P \in \left[0.002, 0.009\right]$. We find that the values of $A_{\parallel} /  k_{\rm B} = 0.33$ K, $P/k_{\rm B} = 0.009$ K work best for our data. This corresponding contribution is represented in green in Figure \ref{fig:Ccontrib}.

Above $10$~K, the lattice degrees of freedom start to contribute significantly. Fitting the data to a Debye model would be difficult because of the ruthenium magnetic transition at $T_{\rm N} = 95$ K. So as a better approximation we chose to measure the specific heat of the non magnetic analogue \LRO : it has the same pyrochlore lattice and therefore we expect a similar phonon contribution. However the \LRO\ measurements need to be rescaled to account for the different molar mass. The lattice specific heat is a function of the ratio $T/\Theta_D$, where $\Theta_D$ is the Debye temperature which varies like the inverse square root of the molar mass. Thus
\begin{equation}
    C_{L,\, \text{HRO}} = C_{L,\, \text{LRO}} \left( T \times \frac{\Theta_{D,\, \text{LRO}}}{\Theta_{D,\, \text{HRO}}} \right)  = C_{L,\, \text{LRO}} \left( T \times \sqrt{\frac{M_{\text{HRO}}}{M_{\text{LRO}}}} \right)
\end{equation}
The lattice contribution is shown with blue points in Figure \ref{fig:Ccontrib}.

\section{AC susceptibility analysis}
We measured the AC susceptibility using a custom noise susceptometer with a very large bandwidth ($1$~mHz - $10$~kHz). The out of phase susceptibility $\chi''$ shows two bumps, which can be qualitatively associated to two time scales: one dominates below the transition at frequencies below $100$~Hz, and another takes over above the transition at frequencies above $1000$~Hz.
\begin{figure}[htbp]
        \centering
        \includegraphics[width=0.83\linewidth]{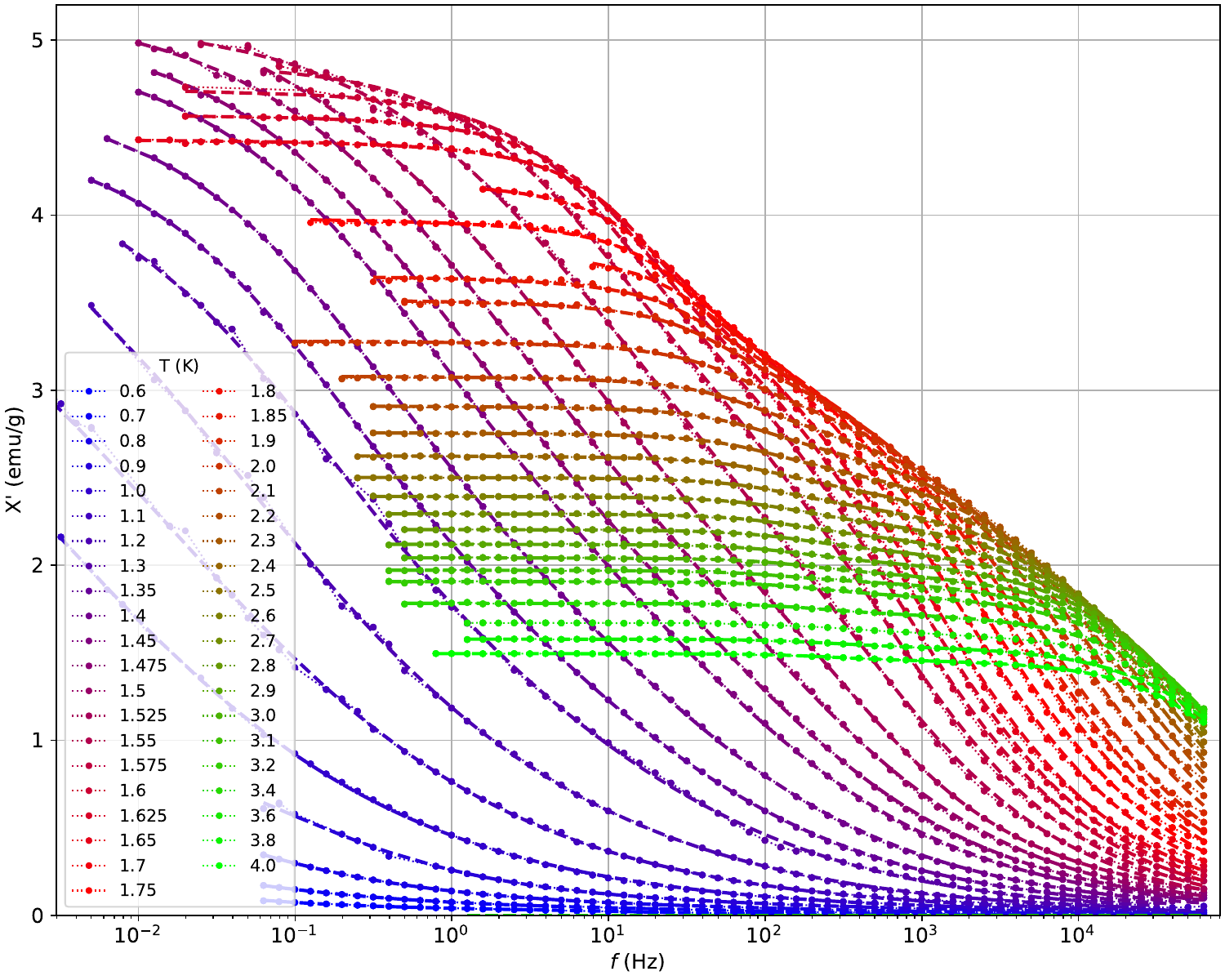}\\
        \includegraphics[width=0.83\linewidth]{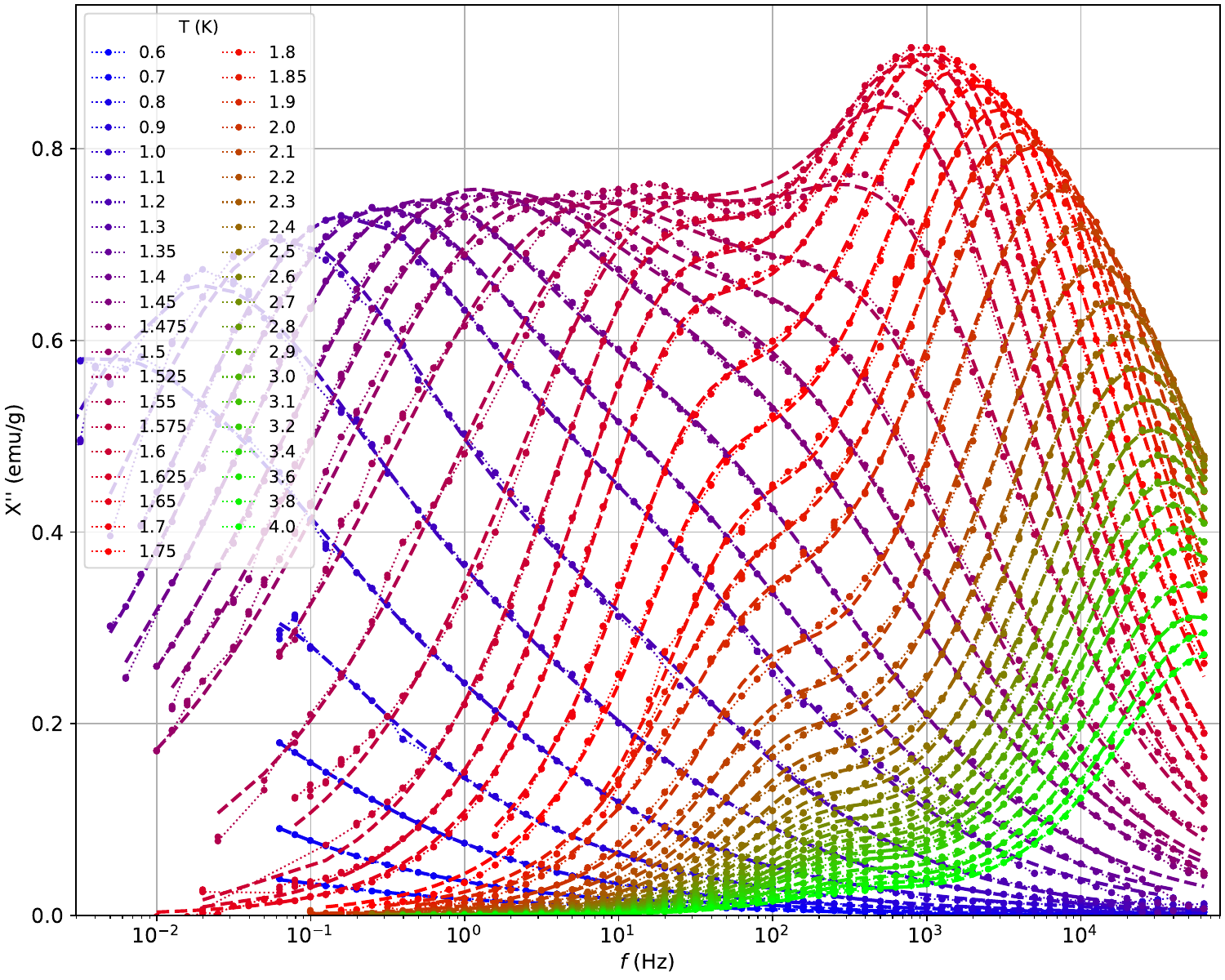}
        \caption{(Top) In phase $\chi'$ and (bottom) and out of phase $\chi''$ susceptibilities measured on the noise magnetometer. The data points are the dots linked by thin dotted lines. The thick dashed lines are fits with equation \ref{fit_chi}.}
        \label{fig:XacNoise}
\end{figure}

Therefore, we elected to fit the data using a model with two relaxation times, each contributing a certain fraction of the total signal. We took into account a symmetric spread of relaxation times through a generalized Debye model:
\begin{equation}
    \chi(f) = \chi_S + (\chi_T - \chi_S) \left( \frac{\eta}{1 + (2 \mathrm{i} \pi f \tau_l)^{1-\alpha_l}} + \frac{1 - \eta}{1 + (2 \mathrm{i} \pi f \tau_s)^{1-\alpha_s}} \right)
    \label{fit_chi}
\end{equation}
where $\chi_T$ is the DC susceptibility, $\chi_S$ is the high frequency limit called adiabatic susceptibility. The two times $\tau_l$ and $\tau_s$ are the characteristic times for the long (or slow) and short (or fast) processes respectively, with spread parameters $\alpha_l$ and $\alpha_s$ and relative amplitude $\eta$. $\alpha = 0$ corresponds to the single time Debye model. 

The fits were performed simultaneously on $\chi'$ and $\chi''$. In order to constrain the fit, the times scales were fitted on the $\chi''$ data, then fixed to fit the rest of the parameters on the $\chi'$ data. The $\alpha$ parameters were also set to $0$ if the amplitude of the related mode is too small. The fits are represented in Figure \ref{fig:XacNoise} by dashed lines. They could not be performed accurately below $1$~K because the relaxation times become longer than $10^3$ seconds.

\begin{figure}[htb]
    \centering
    \includegraphics[width=0.4\linewidth]{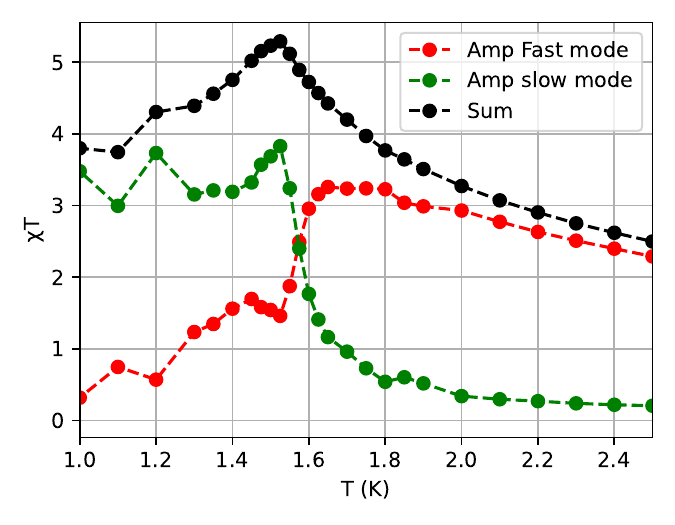}
    \includegraphics[width=0.4\linewidth]{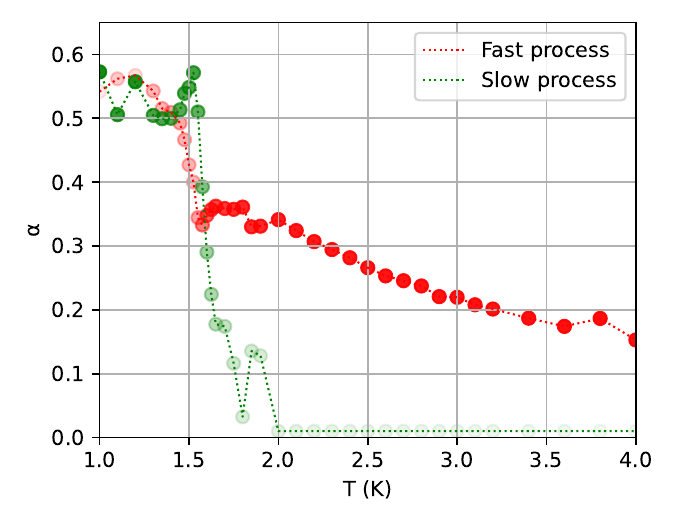}
    \caption{(left) DC susceptibility of each mode in red and green, and their sum in black. (right) Spread of relaxation times $\alpha$ of each mode. The transparency represents the relative amplitude of the modes.}
    \label{fig:tauAlphaChiFit}
\end{figure}
In Figure \ref{fig:tauAlphaChiFit} we show the parameters of the two modes estimated by the fits. The estimated DC susceptibilities for each mode as well as their sum are shown on the left. The slow mode arises very abruptly at the transition, while the fast mode only gets gradually suppressed below the transition. On the right we show the coefficient $\alpha$, which characterizes the spread of relaxation times for each mode. It increases for both modes as the temperature is lowered through the transition; however the fast mode seems to have much broader distribution of relaxation times above the transition than the slow mode.

\section{Inelastic neutron scattering}
Two series of experiments were performed at LET@ISIS, in order to detect possible low energy excitations in inelastic neutron scattering measurements. The first set of measurements was performed with four incident energies, $E_i=2.2$, 3.7, 7.52 and 22.78 meV and at temperatures of 1.5, 20 and 100 K. 

\begin{figure}[htb]
    \centering
    \includegraphics[width=8.3cm]{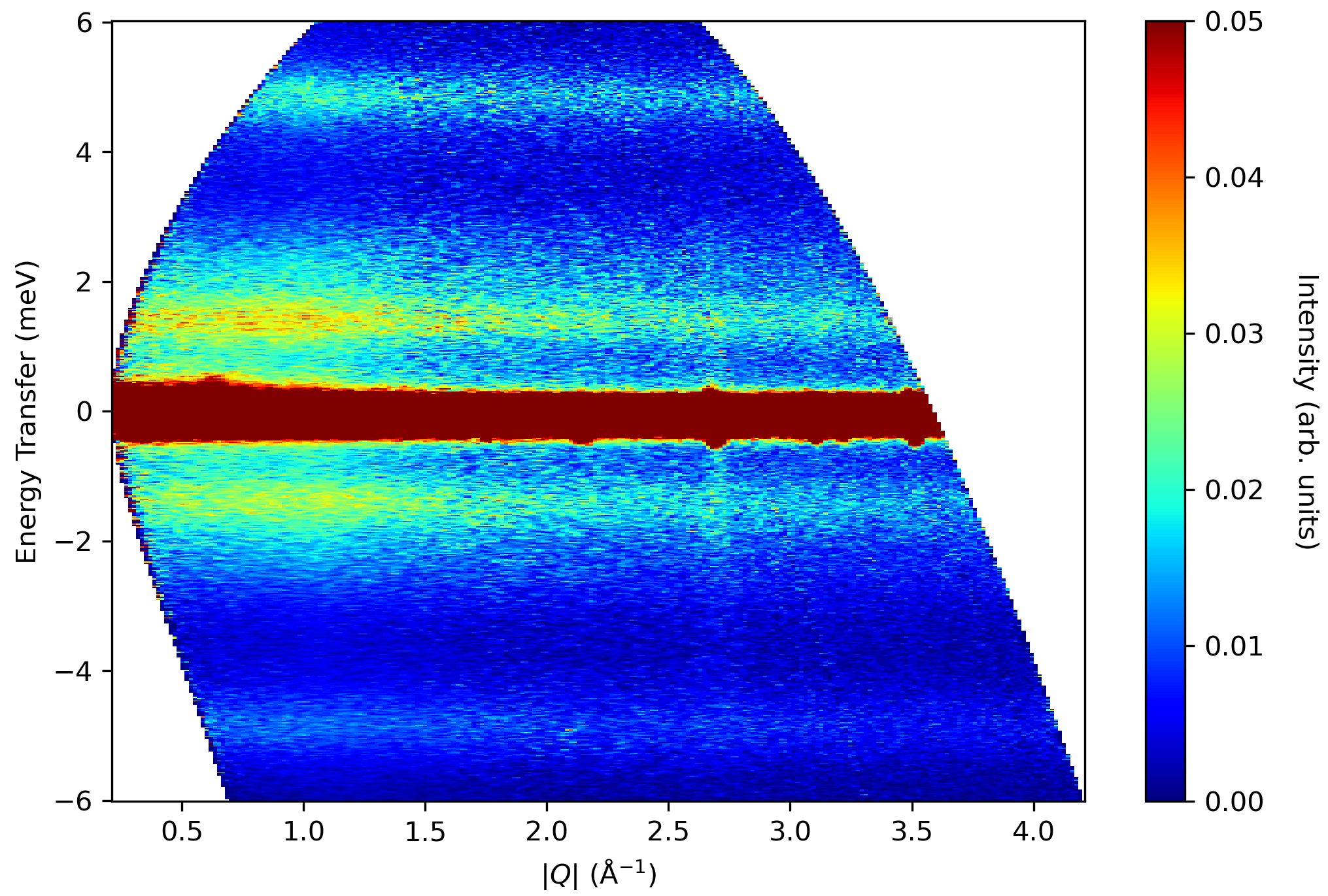}
     \includegraphics[width=8.3cm]{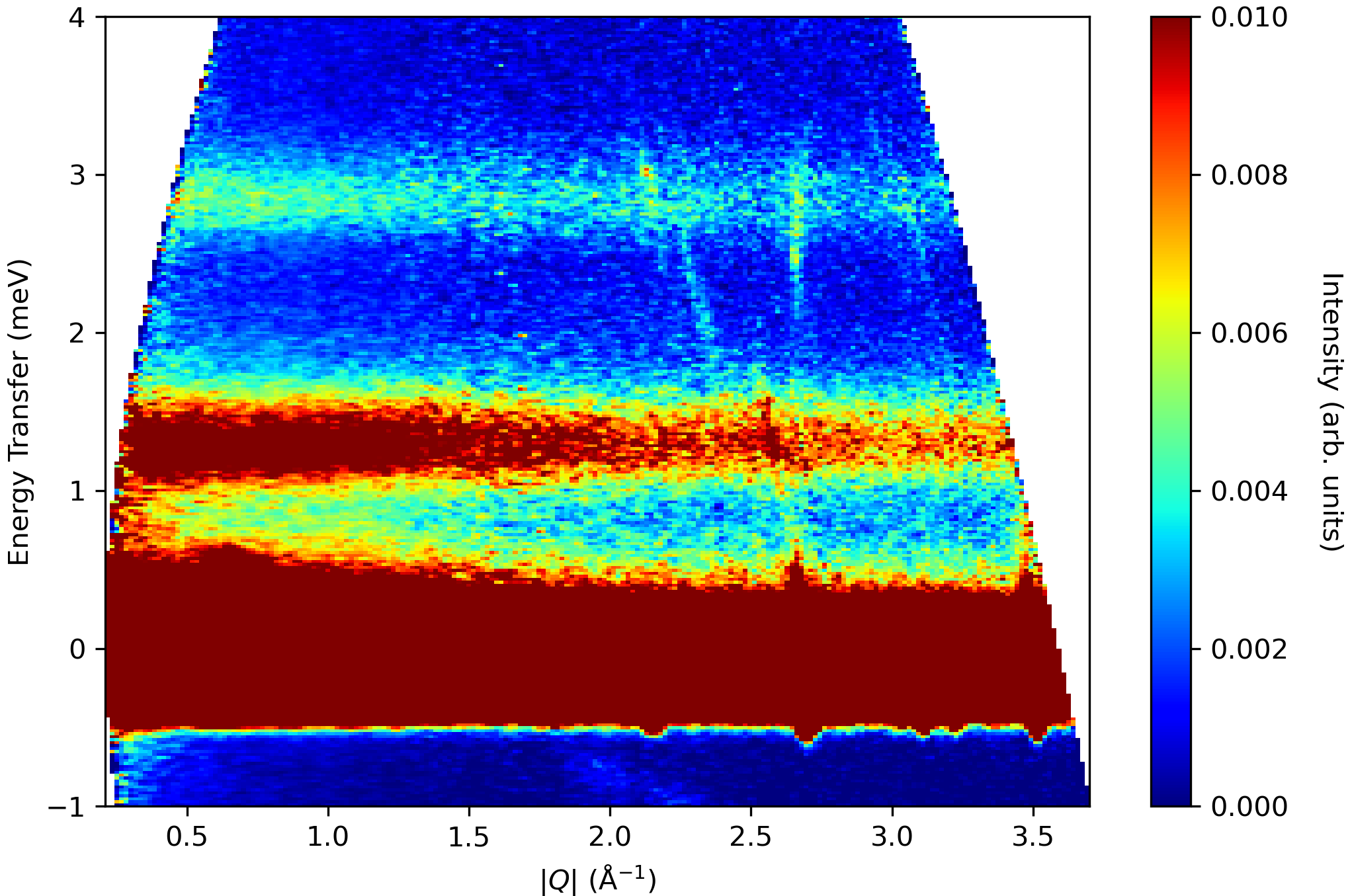}
    \caption{$S(q,E)$ measured with $E_i=7.52$ meV at 100 K (left) and 1.5 K (right). The sharp branches emerging from 2.5 \AA$^{-1}$ are spurious signals. }
    \label{fig:Sqw}
\end{figure}

At 100 K, these measurements could reveal two non-dispersive excitations centered around positive energy transfers 1.5 and 4.8 meV (See Figure \ref{fig:Sqw}(left)). Similar excitations were observed in reference \onlinecite{Wiebe2004}. They can be interpreted as transitions between the excited Ho$^{3+}$ crystal electric field states, which are observed between 20 and 28 meV in \HTO\ \cite{Rosenkranz2000} and are thus thermally populated at the measured temperature of 100 K. 

In Ref. \cite{Wiebe2004}, the authors proposed a CEF scheme with levels between 10 and 20 meV. In our data, no signal is observed between 5 and 18 meV, apart from a phonon around 10 meV, whose intensity is stronger at large $Q$. We thus deduce that the crystal field levels are located above 18 meV, at similar energies as other Ho pyrochlores \cite{Rosenkranz2000, Lefrancois2017, Gaudet2018}.  Our higher incident energy - 22.78 meV - was not large enough to observe these levels, and further measurements are needed to confirm their position.

\begin{figure}[htb]
    \centering
    \includegraphics[height=5cm]{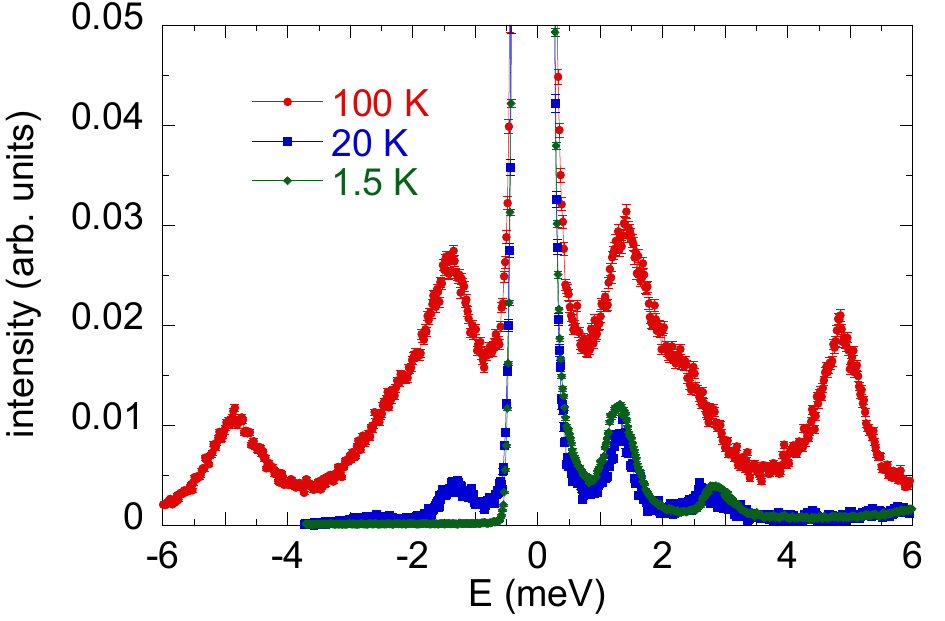}
     \includegraphics[height=5cm]{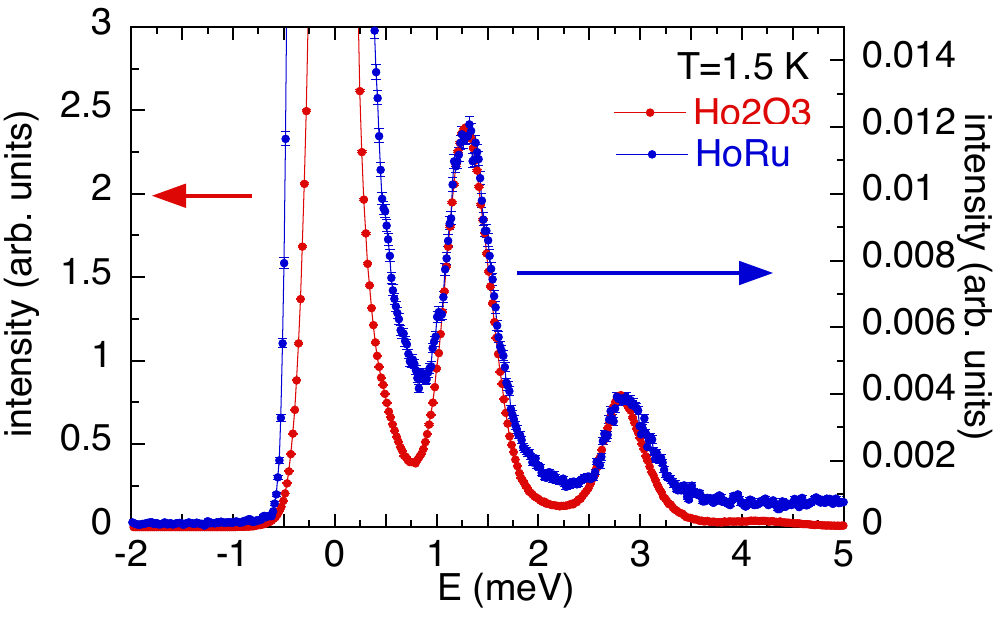}
    \caption{Neutron intensity vs energy transfer obtained from the integration of $S(Q,E)$  between $Q=1.05$ and 1.5 \AA$^{-1}$ with $E_i=7.52$ meV. left: for \HRO\ at 100, 20 and 1.5 K; right: for \HRO\ and Ho$_2$O$_3$ at 1.5 K.}
    \label{fig:Qcut}
\end{figure}

When decreasing the temperature, the signal at negative energy is strongly suppressed and the mode at 4.8 meV disappears (See Figure  \ref{fig:Qcut}(left)), which is consistent with the interpretation in terms of inter CEF level transitions. At 1.5 K, flat modes are nevertheless present at energies of 1.4 and 2.7 meV (See Figures \ref{fig:Sqw}(right)). The origin of these well defined excitations is puzzling, given the non-Kramers nature of the Ho$^{3+}$ ion, and the composition of its ground state doublet wave function in pyrochlore oxide series, which is essentially made of $|\pm 8 \rangle$ and $|\pm 5 \rangle$ components. Indeed transitions between the two states of the doublet cannot be observed in inelastic neutron scattering due to the $\Delta m=0, \pm 1$ neutron selection rule. 

Knowing that an impurity phase of Ho$_2$O$_3$ is present in our sample, we have thus measured a powder sample of Ho$_2$O$_3$ in a second set of experiments, at the same temperatures, and with an incoming energy of 7.52 meV. These experiments could confirm that the low energy signal does come from Ho$_2$O$_3$ (See Figure \ref{fig:Qcut}(right)). Therefore, no clear signal due to \HRO\ is present at these energies. It would be of interest to probe the system at much lower energies to see if transverse terms can generate an inelastic signal. 

\FloatBarrier

\section{Model Hamiltonian}
We have performed preliminary Monte Carlo calculations with the simplest Hamiltonian beyond the molecular field approach which can account for the $\Gamma_5$ ruthenium ordering and the spin ice correlations between the holmium moments. It includes a coupling between the Ho and Ru magnetic moments along the local $z$ directions. The Hamiltonian writes, in the local frame of the magnetic ions: 
\begin{equation}
\begin{aligned}
{\cal H}={\cal H}_{\rm Ho-Ho} + {\cal H}_{\rm Ru-Ru} + {\cal H}_{Ho-Ru} 
\qquad {\rm with} \qquad & {\cal H}_{\rm Ho-Ho} = \sum_{\langle i,j \rangle} J^{\rm Ho-Ho}_{zz}S^z_i S^z_j \\
&  {\cal H}_{\rm Ru-Ru}= - \sum_{\langle i,j \rangle} J^{\rm Ru-Ru}_{\pm}\left(S^+_i S^-_j+S^-_i S^+_j \right) \\
& {\cal H}_{\rm Ho-Ru}=\sum_{\langle i \in {\rm Ho},j \in {\rm Ru} \rangle} J^{\rm Ho-Ru}_{zz}S^z_i S^z_j
\label{hamiltonian}
\end{aligned}
\end{equation}
where the sums act on nearest neighbors. 

Calculations were performed on $N\times N \times N$ unit cells, with $N=4,8$, with  $J^{\rm Ho-Ho}_{zz}/J^{\rm Ru-Ru}_{\pm}=0.02$ to mimic the ratio between the observed $T_{\rm N Ru}=95$ K and the range of nearest neighbor effective Ho-Ho interactions (about 2 K). For  $J^{\rm Ho-Ru}_{zz}/J^{\rm Ru-Ru}_{\pm}>0.2$, a transition of the Ho magnetic moments towards a fully saturated ordered spin ice state is observed, at $T \sim J^{\rm Ho-Ho}_{zz}$. The transition temperature decreases with $J^{\rm Ho-Ru}_{zz}$.

When decreasing further this ratio, we were not able to observe a transition anymore. \\
Actually, when $J^{\rm Ho-Ru}_{zz}/J^{\rm Ru-Ru}_{\pm}<0.2$, the transition would occur below the temperature at which the Ho-Ho spin ice correlations establish. Because we have used a simple algorithm, which is not ergodic, we believe that for these ratio, the system remains blocked in its spin ice configuration, as was shown for the pure spin ice system, and is not able to reach its ground state anymore. It would be necessary to implement a loop algorithm to check if the ordering transition persists for these ratio.

Although they do not allow to obtain the observed ground state of \HRO, these preliminary calculations validate the fact that the interplay between Ho and Ru magnetic moments can induce a phase transition for the Ho moments.

\section{Ferromagnetism along $[111]$ with repulsive cubic field}

The saturated ferromagnetic state of spin ice is a $q=0$ state with magnetisation along any one of the $[100]$ directions. Each spin has a projection along the cubic direction of $\frac{1}{\sqrt{3}}$ so that, in the $[100]$ ground state, the vector moment per spin is $\vec M^{max} = \frac{1}{\sqrt{3}} \hat{z}$ where $\hat{z}$ is along one of the six possible directions and the moment per spin is taken to be unity.

This should be contrasted with a fragmented ferromagnetic phase with moment along (or against) one of the body centred diagonal directions
\begin{eqnarray} 
\vec d_{1}&=&\frac{1}{\sqrt{3}}[1,1,1], \; \vec d_2=\frac{1}{\sqrt{3}}[1,-1,-1]\nonumber \\
 \vec d_3=&=&\frac{1}{\sqrt{3}}[-1,-1,1], \; \vec d_4=\frac{1}{\sqrt{3}}[-1,1,-1].
\end{eqnarray}
For $\vec M$ lying along one of these axes $\hat{d}$, the maximum total magnetisation of an up tetrahedron (which we take to be the unit cell of the pyrochlore lattice) is $M_{tet}=1+\frac{1}{3}+\frac{1}{3}-\frac{1}{3}=\frac{4}{3}$ so that $\vec M^{max}=\frac{1}{4}M_{tet}\hat{d}=\frac{1}{3}\hat{d}$. Making reference to the fragmentation picture we refer to these two states as saturated (the $[100]$ state) and fragmented (the $[111]$ state).  

As these two ground states already have cubic symmetry, a self-consistent field term could separate them allowing for the fragmented state to be the preferred low energy state. 

The ferromagnetic states would be selected through the generation of a perturbation that lifts the spin ice manifold (quasi)-degeneracy. One possibility is the $J_{3a}$ third neighbour coupling as defined in reference \onlinecite{Henelius2016}.  To this we add a quartic term depending on a self-consistent field, $\vec h_{eff}= \vec M$ whose sign will define which of the ferromagnetic states are selected. A perturbation to the spin ice Hamiltonian could take the form
\begin{equation}
H_1=-J_{3a}\sum_{ij} \vec S_i.\vec S_j + \frac{v_0}{4}\sum_{I,N_I} \left(\sum_{j=1,4}\vec S_i.\vec h_{eff}\right)^2,
\end{equation}
where the first sum runs over all third neighbour pairs coupled by $J_{3a}$ \cite{Henelius2016} and the second runs over all spins separated into $N_I$ unit cells ($N=4N_I$). 
%The idea is that the self-consistent term with $v_0$ defined here to be positive, comes from a pot pourrie of supplementary terms that, because of the long range interactions, have the cubic symmetry of the lattice. I took a quartic term because we need a term of different form from that of the exchange coupling. A quadratic coupling of the form $v_0\sum_i \vec S_i.\vec M$ looks like a mean field treatment of the exchange term, so that, at the mean field level one would be able to absorb $v_0$ into an effective $J$. Hence, the extra term must appear with a higher power of the order parameter than quadratic. I summed over a unit cell as this is the smallest repeating unit of spins in the $q=0$ state.  %However, taking a quadratic coupling does not actually change the qualitative result. 

The ground state energies can be estimated for the two states as a function of $v_0$. 
%For an up-tetrahedron, the four spins have a total of $24$, $3^{rd}$ neighbour bonds, but each will be counted twice as the coupling is between different unit cells. 

\begin{itemize}

\item In the saturated state, all $3^{rd}$ neighbour pairs lie parallel to each other, giving an exchange energy of $-12J_{3a}$ per cell, or $-3J_{3a}$ per spin. The self-consistent field is $h_{eff}= \frac{1}{\sqrt{3}} \hat{z}$ so that the perturbation per spin is
\begin{equation}
\epsilon_s=-3J_{3a}+\frac{4v_0}{9}.
\end{equation}

\item In the fragmented state, the six $3^{rd}$ neighbour bonds of the apical spins are satisfied, but the $18$ others remain frustrated. However, their total contribution will not be zero because of the 2-in-1-out ice rules. This broken $Z_2$ symmetry means that the mean of the third neighbour correlation functions for spins in the kagome plane will be $\left<\vec S_i.\vec S_j\right>=\frac{1}{9}$
%. This means that the energy contribution from the third neighbour coupling should be $-6J_{3a}-2J_{3a}=-8J_{3a}$ per tetrahedron, 
leading to $-4J_{3a}$ per unit cell and $-J_{3a}$ per spin. 
The self-consistent field is $h_{eff}=\frac{1}{3}\hat{d}$ so that the perturbation energy per spin is $\frac{v_0}{4}\left(\frac{1}{3}+\frac{1}{9}\right)^2=v_0\left(\frac{4}{81}\right)$ and the total energy per spin is
\begin{equation}
\epsilon_f=-J_{3a}+\frac{4v_0}{81}.
\end{equation}

\end{itemize}

Putting these two energies equal we find that the fragmented phase becomes stable for 
\begin{equation}
v_0=\frac{81}{16}J_{3a}\;, \epsilon_f=-\frac{3}{4}J_{3a}.
\end{equation}
%
%and
%
%\begin{equation}
%\epsilon_f=-\frac{3}{4}J_{3a}.
%\end{equation}
%

\vspace{1cm}
%\bibliography{biblio_supmat}

\end{document}